\documentclass{article}

\usepackage[final]{nips_2017}

\usepackage[utf8]{inputenc} 
\usepackage[T1]{fontenc}    
\usepackage{hyperref}       
\usepackage{url}            
\usepackage{booktabs}       
\usepackage{amsfonts}       
\usepackage{nicefrac}       
\usepackage{microtype}      
\usepackage{graphicx}

\title{Generative Models for Spear Phishing Posts on Social Media}

\author{
  John Seymour \\
  Salesforce \\
  \texttt{john.seymour@salesforce.com} \\
  \And
  Philip Tully \\
  ZeroFOX \\
  \texttt{ptully@zerofox.com} \\
}

\begin{document}

\maketitle

\begin{abstract}
Historically, machine learning in computer security has prioritized defense: think intrusion detection systems, malware classification and botnet traffic identification. Offense can benefit from data just as well. Social networks, with their access to extensive personal data, bot-friendly APIs, colloquial syntax, and prevalence of shortened links, are the perfect venues for spreading machine-generated malicious content. We aim to discover what capabilities an adversary might utilize in such a domain. We present a long short-term memory (LSTM) neural network that learns to socially engineer specific users into clicking on deceptive URLs. The model is trained with word vector representations of social media posts, and in order to make a click-through more likely, it is dynamically seeded with topics extracted from the target’s timeline. We augment the model with clustering to triage high value targets based on their level of social engagement, and measure success of the LSTM’s phishing expedition using click-rates of IP-tracked links. We achieve state of the art success rates, tripling those of historic email attack campaigns, and outperform humans manually performing the same task.

\end{abstract}

\section{Introduction}

Historically, the security community has used machine learning in a defensive manner, for example in classifying malicious binaries or finding anomalous network traffic. Even now, startups continue to spring up advertising novel techniques for detecting inbound threats. But much of computer security is dedicated to identifying critical weaknesses and vulnerabilities through a simulated offense. While tools such as Metasploit [1] exist for automating Red-Team activities, there has been little exploration into how machine learning might be used in such a manner. 

Social engineering is when an attacker “influences a person to take an action that may or may not be in their best interest” [11]. One of the most prevalent types of social engineering is phishing, or when an attacker directs a user to give them privileged information while pretending to be someone the user trusts. The user then gives this privileged information to the attacker directly. Phishing strategies range from the shotgun approach, where messages are sent en masse to targets and only a few are expected to bite, to the highly targeted “spear-phishing” approach, where messages have increased probability of success from manual curation of personal information which is then used to gain the target’s trust. Phishing attacks span across a variety of platforms, but are especially prevalent on social media, with their ever-expanding user bases, high usage statistics, and strong incentives for disclosing personal data.

In computer security, Natural Language Processing (NLP) has been successfully used for many applications, one of significance being spam detection [3]. However, phishing is an interesting use case, as machines have been successfully fooling humans dating back to the ELIZA chatbot in 1966 [2]. Phishing is especially amenable to the NLP approach because recurring patterns of text can be utilized both to identify topics the target is interested in and to generate sentences the target might respond to. If one could automatically churn through unorganized personal data and generate a personalized phishing message, they could combine the shotgun approach of phishing campaigns with the specificity and success of targeted spear phishing.

\section{Related Work}

Several studies report the effectiveness of phishing and spear phishing attacks. An early study found between 7-11\% of users will click on a malicious link, depending on whether the sender includes their name in the message [8]. More recently, the 2017 Verizon Data Breach Investigation Report estimates the number of users in breaches who are fooled by a phishing email to be 7.3\% [10]. Google reports that 13.78\% of users will enter their credentials to a phishing site, based on their internal Gmail and Google SafeBrowsing data [9].

For automating social engineering attacks, the Social-Engineer Toolkit is the gold standard [4]. It is an open-source, Python toolkit, which generates the payloads for social engineering: the files that a victim opens and either enters their credentials or runs a malicious program. For example, some files that the toolkit can generate are web pages, PDFs, or Java Applets.

However, the toolkit only automates the final stage in the pipeline. Users of the toolkit must still have their victims execute the generated programs with a “front-end” delivery service. They must still, for example, write an email which gets the user to download the payload, using one of the two phishing strategies.

A proof of concept for automating the delivery mechanism, Honey-Phish, generated an email which attempted to trick scammers into clicking a link [5]. Since a corpus of phishing emails was unavailable, and believing most scammers to be interested in financial terminology, Honey-Phish used a Hidden Markov Model [6] trained on Reddit posts from /r/personalfinance. However, the model-generated English was noticeably different than what a human would write, and only 2 phishers responded out of 41 phishing emails generated.

We make this approach viable by switching the environment: we post on social media instead of sending an email. This allows us to scale both on the number of targets and on accuracy by tailoring the messages using personal data. Furthermore, social media users readily accept broken English because of platforms’ real-time nature that emphasizes brevity, informal culture, and character limits.

\section{Methods}

Our tool uses two different machine learning models in order to optimize click-through rates. Our first model triages the users into clusters for value and engagement. It takes as input a streaming API to collect a large number of users. Our second model generates text and a simulated “phishing” URL using the timeline of the target, which is then posted by a social media account that we crafted with embellished profile fields and legitimate interactions with other users on the platform in order to seem more believable. We consider a click from one of these URLs to be a success, and compare resulting metrics with controlled experiments and statistics from previous literature.

If an attacker indiscriminately spammed lots of users with phishing links, the dummy accounts would quickly be discovered and terminated for Terms of Service (ToS) violations. Therefore, we triage users to determine which ones are either more likely to be phished or provide exceptional value. We cluster users into groups based on their overall user data, such as whether they have changed default settings, follower interactions, and engagement metrics. 

We use KMeans with k=3 clusters to triage users. We compared this clustering method with DBSCAN, Birch, Affinity Clustering, and Spectral Clustering, using a randomized grid search to determine hyperparameter settings for each. The selected algorithm achieves a Silhouette Score of 0.71 with a reasonable cluster structure. The users in the cluster selected displayed relevant features toward phishing susceptibility, such as high follower count, high post count, descriptive bio text including job descriptions like “CEO” and company names, creation dates in the distant past, and many changes from the default account settings.

If the user fits into the cluster that we’ve empirically identified as likely to lead to a successful phish, we additionally collect their posting history in order to generate the targeted message. Our automated posts are made up of three separate components: a target username, the machine-generated content, and lastly a goo.gl shortened URL. 

\begin{figure}[h]
  \centering
  \fbox{\includegraphics{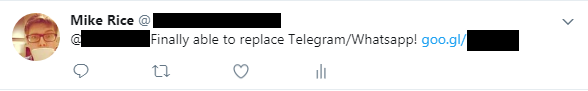}}
  
  \caption{Example of a machine-generated post.}
\end{figure}

To restrict visibility of our posts to only their intended recipients and avoid unwanted attention, we make use of a special “Reply” functionality. Replies are visible only to the author, the specified username, and any followers the two accounts share in common with each other, and consist of any posts beginning with the symbol “@”. We ensure our dummy account doesn’t follow any other users, and so our posts are only visible to us and to our targets.

The content of the post is generated by a long short-term memory (LSTM) model. We pre-train a model using the Twitter GloVe embeddings [12] to attempt to control for language patterns in social media. We attempted several methods for extracting a topic from the user’s posts to use as a seed for the generated text, but we found a naive method based on dictionary frequency and stop words to be as effective as more complex methods. Alternatively, the tool can be configured to train a Hidden Markov Model (HMM) directly on the user’s posts. In testing, we found HMMs to overfit for users with a small number of posts, but have the added benefit of adapting to users who post in other languages.

To simulate the “phishing” link, we use Google’s URL shortening service [13] which uniquely shortens input URLs (e.g. \url{www.google.com} is shortened to \url{https://goo.gl/yXcpef}). The goo.gl URL is convenient, because it is hosted by a trustworthy brand and HTTPS-secured, decreasing a target’s suspicion. The service also provides metrics for each shortened URL, including whether the URL has been retrieved and by whom. These metrics can be queried through an API, allowing programmatic gathering of whether each link was clicked. To avoid harm to targets we use \url{www.google.com} as our “phishing” destination, and to obtain accurate click-through rates, we append the target’s username as a parameter, (e.g. \url{https://www.google.com/?screen_name=[TARGET]}).

In order to increase the effectiveness of the posts, we schedule this post for when the target is likely to be online to receive it. To do so, we extract the times of the posts from the user’s history, generate a histogram of these post timings binned by hour, select the hour when the user was most active, and post the simulated phishing message during a random minute of that selected hour.

\begin{figure}[h]
  \centering
  \fbox{\includegraphics{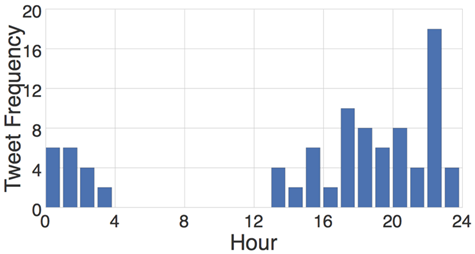}}
  \caption{Example histogram of a user’s number of social media posts per hour. The model 
selects a random minute during hour 22 to post the phishing message.}
\end{figure}

\section{Experimentation}

We validate our models with two different methods. Our first test measures the effective click-through rate of the entire tool including post scheduling. Because this additional feature adds confounding variables for throughput, we also run the tool in a controlled environment without post scheduling. During the second test, we competed with a human performing the same task.

Our test for measuring the effective click-through rate was structured as follows. We sent 90 “phishing” posts to users of a common hashtag, including scheduling. Because there are multiple causes for latency (both the time between generation of the text and posting, as well as the time between posting and the target reading) we waited two days before measuring the clickthrough rate.

All of the goo.gl reports showed the links had been visited, but we spent some effort to discern whether a human visited the link or whether the link had been visited by web crawlers. For example, social media sites themselves will crawl links posted on their site to determine whether the link is on a public blacklist. From the reports, 30\% of our clickthroughs came from a t.co referrer, which we take to be actual humans clicking the link through the official web application. Also from the reports, an additional 36\% came from unique countries where such bots are unlikely to tunnel through. We thus report a clickthrough rate between 30\% and 66\%.

We also wished to obtain a metric for how many targets per unit time such a system could achieve. We decided to run a single instance of the tool over two hours to measure this quantity. A human competitor attempted to perform the same task in the same timeframe, using a strategy of pre-built messages targeting a specific group of users. We measured our click-through rate as in experiment 1, including waiting two days. Our results are shown in Table 1. We note that the largest bottleneck for our system is the rate-limiting imposed by the social network for its API, and that removing the scheduling capability drastically reduces our clickthrough rate.

\begin{table}[t]
  \caption{Throughput of human and tool over two hours}
  \label{bakeoff}
  \centering
  \begin{tabular}{lll}
    \toprule
    Metric & Human & Tool\\
    \midrule
    Total Targets & 129 & 819\\
    Posts/minute & 1.075 & 6.85\\
    Maximum Click-throughs & 49 & 275\\
    \bottomrule
  \end{tabular}
\end{table}

\section{Conclusion}

Though large-scale phishing campaigns tend to have very low compromise rates, they persist because the few examples that do succeed lead to a high return on investment. We found that our automated spear phishing framework was more successful than the 5-14\% previously reported in large-scale campaigns [7-10]. We attribute our results to the unique risks associated with social media and our ability to leverage data science to target vulnerable users with a highly personalized messages.

This work marks an advance in offensive capabilities through automation of a traditionally manual process using machine learning techniques. Our approach is predicated on the fact that social media is rapidly emerging as an easy target for phishing and social engineering attacks with its low bar for admissible messages, its community tolerance of convenience services like shortened links, its effective API, and its pervasive culture of overexposing personal information. A specific benefit of social media compared to traditional media is the length of messages: because short messages are the norm, messages produced by the model have decreased probability of grammatical error.

There are existing frameworks such as the Social-Engineer Toolkit that automate the payload of the phishing process, but none that tailor the phishing message to the target. We close this gap and enable penetration testers to address larger groups of targets while not compromising the quality of the spear phishing message. Due to the democratization of machine learning, we believe this capability to be accessible to adversaries in the near future, and we present this research in order to contribute to the development of suitable countermeasures.

\section*{References}

\small

[1] Maynor, David. Metasploit toolkit for penetration testing, exploit development, and vulnerability research. Elsevier, 2011.

[2] Weizenbaum, Joseph. "ELIZA—a computer program for the study of natural language communication between man and machine." Communications of the ACM 9.1 (1966): 36-45.

[3] Sahami, Mehran, Dumais, Susan, Heckerman, David, and Eric Horvitz. "A Bayesian approach to filtering junk e-mail." Learning for Text Categorization: Papers from the 1998 workshop. Vol. 62. 1998.

[4] Pavković, Nikola, and Luka Perkov. "Social Engineering Toolkit—A systematic approach to social engineering." MIPRO, 2011 Proceedings of the 34th International Convention. IEEE, 2011. https://www.trustedsec.com/social-engineer-toolkit/

[5] Gallagher, Robbie, “Where Do the Phishers Live? Collecting Phishers’ Geographic Locations from Automated Honeypots”, 2016 ShmooCon, https://bitbucket.org/rgallagh/honey-phish

[6] Markov, Andrey A. "Extension of the limit theorems of probability theory to a sum of variables connected in a chain". reprinted in Appendix B of: R. Howard. Dynamic Probabilistic Systems, volume 1: Markov Chains. John Wiley and Sons, 1971.

[7] Thompson, Steven C. "Phight Phraud." Journal of Accountancy 201.2 (2006): 43.

[8] Jakobsson, Markus, and Jacob Ratkiewicz. "Designing ethical phishing experiments: a study of (ROT13) rOnl query features." Proceedings of the 15th international conference on World Wide Web. ACM, 2006.

[9] Bursztein, Elie, et al. "Handcrafted fraud and extortion: Manual account hijacking in the wild." Proceedings of the 2014 Conference on Internet Measurement Conference. ACM, 2014.

[10] “The Verizon DBIR.” Verizon Enterprise Solutions, www.verizonenterprise.com/verizon-insights-lab/dbir/.

[11] “The Official Social Engineering Portal.” Security Through Education, www.social-engineer.org/.

[12] Jeffrey Pennington, Richard Socher, and Christopher D. Manning. 2014. GloVe: Global Vectors for Word Representation.

[13] “Google URL Shortener.” Google, https://goo.gl/.

\end{document}